\documentclass[pdftex,iop,apjl]{emulateapj} 

\usepackage{longtable}
\usepackage{epstopdf}  
\usepackage{ifthen}
\usepackage{rotating}
\usepackage{subfigure} 
\usepackage{booktabs}

\usepackage{appendix}
\usepackage{color}

\submitted{Accepted for publication in {\it The Astrophysical Journal Letters}, 2017 November 25}
\shorttitle{{\sc On the rotation period and shape of 1I/`Oumuamua}}
\shortauthors{{\sc Knight et al.}}

\def\icarus{Icarus}

\begin{document}
\bibliographystyle{apj}

\title{On the rotation period and shape of the hyperbolic asteroid 1I/`Oumuamua (2017 U1) from its lightcurve}

\author{Matthew M. Knight\altaffilmark{1,2}, Silvia Protopapa\altaffilmark{2}, Michael S.~P. Kelley\altaffilmark{2}, Tony L. Farnham\altaffilmark{2}, James M. Bauer\altaffilmark{2}, Dennis Bodewits\altaffilmark{2}, Lori M. Feaga\altaffilmark{2}, Jessica M. Sunshine\altaffilmark{2}}

\altaffiltext{1}{Contacting author: mmk8a@astro.umd.edu.}
\altaffiltext{2}{University of Maryland, Department of Astronomy, 1113 Physical Sciences Complex, Building 415, College Park, MD 20742, USA}

\begin{abstract}
We observed the newly discovered hyperbolic minor planet 1I/`Oumuamua (2017 U1) on 2017 October 30 with Lowell Observatory's 4.3-m Discovery Channel Telescope. From these observations, we derived a partial lightcurve with peak-to-trough amplitude of at least 1.2 mag.  This lightcurve segment rules out rotation periods less than 3 hr and suggests that the period is at least 5 hr. On the assumption that the variability is due to a changing cross section, the axial ratio is at least 3:1. We saw no evidence for a coma or tail in either individual images or in a stacked image having an equivalent exposure time of 9000 s.\end{abstract}


\section{INTRODUCTION}
The discovery of the minor planet 1I/`Oumuamua (2017 U1) was announced on 2017 October 25 \citep{mpec2017u181} and its orbit was definitively found to be hyperbolic ($e=1.1994{\pm}0.0002$; JPL Small-Body Database reference JPL13, 2017 November 13). As the first minor planet of extra-solar origin to be identified in our solar system, it is the subject of intense interest. At discovery, `Oumuamua had already passed perihelion (2017 September 9) and continued to rapidly fade, making the window for measuring its physical properties extremely short. 
	
There was initial confusion over the asteroidal versus cometary nature of `Oumuamua. This was primarily due to its original designation as ``C/2017 U1,'' which was subsequently revised to ``A/2017 U1'' with a note that it was ``inadvertently designated as [a] comet...with no claimed cometary appearance'' \citep{cbet4450a}. Although it was reported that ``a very deep, stacked image, obtained with the Very Large Telescope'' showed it to be ``completely stellar'' \citep{mpec2017u183}, no quantifiable assessment of these observations had been reported at the moment of submitting this manuscript. Further, while astrometry and snapshot magnitudes in a variety of bandpasses have been regularly reported and we are aware of at least three optical spectra \citep{cbet4450b,masiero17,ye17}, no lightcurve measurements, which yield insight into the physical characteristics of the object, had been reported at the time the manuscript was submitted. While this paper was under review, four papers presenting lightcurves were made public \citep{bannister17,bolin17,jewitt17,meech17}. However, we have restricted this analysis to our own imaging and lightcurve observations of our solar system's first known visiting minor planet.

\section{OBSERVATIONS AND REDUCTIONS}
\label{sec:observations}
We observed `Oumuamua on 2017 October 30 from 4:18 to 7:07 UT with Lowell Observatory's 4.3-m Discovery Channel Telescope (DCT). Conditions were photometric and the seeing was stable at 0.7--1.0 arcsec.  `Oumuamua was at a heliocentric distance of 1.48 au, a geocentric distance of 0.57 au, and a phase angle of 24.7$^\circ$.  During our observations, it spanned an airmass range of 1.15--1.43. Our data were collected when `Oumuamua was 28$^\circ$ away from a 72\% illuminated Moon. Observations were made using the Large Monolithic Imager \citep{massey13} which was binned 2$\times$2 on chip, yielding 0.24 arcsec pixel$^{-1}$ resolution across a 12.3$\times$12.3 arcmin field of view. All images were tracked at the object's ephemeris rate ($\sim$40 milliarcsec s$^{-1}$) and were obtained in the {\it VR} filter. The {\it VR} filter is a wide filter having a central wavelength of 6092~{\AA} and full-width half maximum (FWHM) of 1764~{\AA} \citep{massey17}. It was developed for deep surveys since it provides nearly a factor of 2 more throughput than a traditional $R$ filter with only a small increase in the sky background \citep{jewitt96}. We obtained 30 consecutive 300~s observations, dithering every three frames, along with a single short image at the beginning (30~s) and at the end (3~s) of the sequence in which the stars were minimally trailed. All images have been processed by applying our standard image processing procedures (e.g., \citealt{knight15a}) to remove the bias and perform flat-field corrections; we used images of a diffusely illuminated spot on the inside of the dome to construct the flat-field images.

Since the stars trailed by $\sim$50 pixels in each 300~s exposure, necessitating very large photometric apertures, it was not practical to derive a photometric calibration directly from each image. Instead, we determined a zero point for the night by analyzing aperture photometry on 89 field stars in our shortest exposure (3~s, resulting in stars trailed by $\sim$0.5 pix) and then applied an airmass correction (discussed in the next paragraph). We measured instrumental magnitudes using a photometric aperture of radius 8 pixels (the FWHM including trailing was 3.2 pix) and determined sky within an annulus from 20 to 40 pixels. We looked up each star's PanSTARRS (PS1) {\it g'} and {\it r'} magnitudes (denoted {$g'_{P1}$ and $r'_{P1}$; \citealt{flewelling16}), determined the offset from the instrumental magnitude to $r'_{P1}$ for each, and performed a linear regression to the offset as a function of ($g'_{P1} - r'_{P1}$). The resulting color term is small, $-0.062(g'_{P1} - r'_{P1}$), so we assumed a solar color of 0.405 ($B-V$ from \citealt{colina96} converted to the PS1 system using the relationships of \citealt{tonry12}) for converting our instrumental magnitudes to $r'_{P1}$.  The final uncertainty on the zero point is based on the scatter in the offsets, 0.01 mag.  We note that `Oumuamua's color has been reported to be slightly redder than solar (e.g., 10\% per 0.1 $\mu$m$^{-1}$; \citealt{ye17}) but the difference from solar color has a small effect in our absolute calibrations, $\leq$0.01 mag.  We thus obtained absolute calibrations by using the zero point derived above and applying an airmass correction to each image, described next.

We assumed an airmass correction of 0.14 mag airmass$^{-1}$ which was interpolated from typical airmass corrections for the $V$ and $R$ filters at the DCT. We tested the accuracy of the assumed extinction correction by measuring the flux of four reference stars visible in all images using a 40-pixel radius aperture (sufficient to encompass the trailing which was $\sim$50 pixels in length) and a sky annulus with inner and outer radii of 50 and 75 pixels. The stars showed a slight trend of brightening with increasing airmass, indicating that our adopted slope slightly overestimates the true correction. However, since the standard deviation on the instrumental magnitude for each star over the night was $<$0.01 mag, which is acceptable for the purposes of the results discussed in this paper, we did not pursue any additional adjustments to the photometric calibrations. We estimate that the systematic uncertainty is 0.02 mag, which incorporates the uncertainty in the zero point (0.01 mag) and in the airmass correction (0.01 mag).

We measured the brightness of `Oumuamua within a circular aperture of radius 8 pixels (the same size used for absolute calibrations and approximately twice the typical seeing, which was 0.7--1.0 arcsec or 3--4 pixels) centered on the object and used a sky annulus with inner and outer annuli of 15 and 50 pixels, respectively. We explored a variety of aperture and annuli sizes and found consistent results for both the stars and the object, confirming that our analysis was not compromised by the strong moonlight. Fortuitously, no images contained obvious background stars that transited our photometric aperture. Aperture fluxes were converted to $r'_{P1}$ magnitudes using the photometric zero point and extinction correction discussed above. The $r'_{P1}$ band lightcurve is shown in Figure~\ref{fig:rad_prof} and given in Table~\ref{t:data}, with UT times given at the midpoint of each image. The error bars indicate the stochastic uncertainties and do not include the 0.02 mag calibration uncertainty previously discussed, which should not affect interpretation of the lightcurve due to its large amplitude. Consistent with the lightcurve, `Oumuamua was initially easily visible in a single 300~s frame, but faded rapidly and nearly disappeared around 5:30 UT, before brightening until our observing block ended.

\begin{figure}[ht]
  \centering
  \includegraphics[width=60mm]{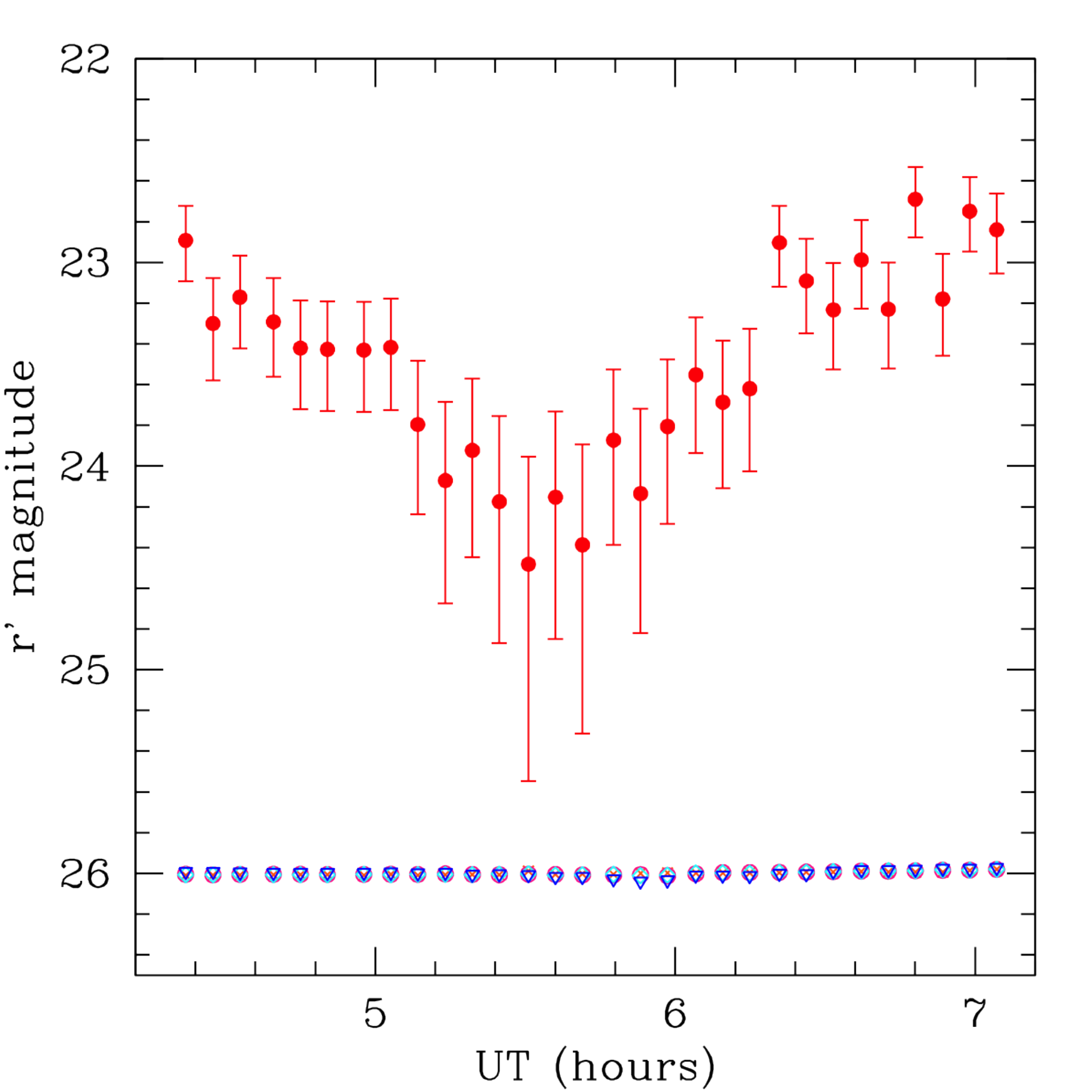}
  \caption[Lightcurve]{Lightcurve of `Oumuamua on 2017 October 30 (red circles) in the PanSTARRS $r'$ system. Four field stars observed in all frames are also plotted, with each star offset so that its average magnitude for the night is $r'_{P1} = 26.0$. Star magnitude uncertainties are not plotted because they are much smaller than the data points. The very flat star magnitudes demonstrate that conditions were stable throughout the night and that the large photometric variations in `Oumuamua's lightcurve are due to the asteroid.}
  \label{fig:lightcurve}
\end{figure}


\begin{deluxetable}{cccc}  
\tabletypesize{\scriptsize}
\tablecolumns{4}
\tablewidth{55mm} 
\setlength{\tabcolsep}{0.05in}
\tablecaption{Photometry on 2017 October 30 UT calibrated to the PanSTARRS1 magnitude system }
\tablehead{   
  \colhead{UT}&
  \colhead{$r'_{PI}$}&
  \colhead{${\sigma}r'_{P1,upper}$}&
  \colhead{${\sigma}r'_{P1,lower}$}
}
\startdata
4.368&22.892&0.169&0.201\\
4.459&23.300&0.223&0.280\\
4.549&23.171&0.205&0.252\\
4.660&23.292&0.216&0.270\\
4.750&23.421&0.235&0.300\\
4.840&23.427&0.236&0.303\\
4.962&23.431&0.236&0.303\\
5.052&23.417&0.239&0.308\\
5.142&23.796&0.312&0.440\\
5.234&24.071&0.386&0.604\\
5.323&23.923&0.352&0.524\\
5.413&24.175&0.420&0.695\\
5.511&24.482&0.527&1.065\\
5.601&24.153&0.421&0.697\\
5.691&24.387&0.492&0.926\\
5.794&23.873&0.348&0.515\\
5.884&24.135&0.416&0.684\\
5.974&23.806&0.330&0.477\\
6.069&23.552&0.283&0.384\\
6.159&23.687&0.303&0.422\\
6.249&23.620&0.294&0.405\\
6.347&22.903&0.181&0.217\\
6.437&23.091&0.208&0.257\\
6.527&23.233&0.230&0.292\\
6.621&22.988&0.195&0.239\\
6.711&23.230&0.230&0.292\\
6.801&22.690&0.159&0.187\\
6.892&23.180&0.222&0.279\\
6.982&22.749&0.168&0.199\\
7.072&22.840&0.179&0.214
\enddata
\label{t:data}
\label{lasttable}
\end{deluxetable}


\newpage
\section{RESULTS AND ANALYSIS}\label{sec:results}
The lightcurve displays a clear minimum at ~5:30 UT. Although there are some hints of flattening at both the beginning and end of our observations, we do not obviously bracket a maximum in either case. We place a lower limit of 1.2 mag on the peak-to-trough range of the lightcurve.  If the lightcurve variability is due to the varying cross-section of a prolate spheroid, then `Oumuamua has an axial ratio of at least 3:1; i.e., it is highly elongated. 

Alternatively, the variability could be caused, at least in part, by large albedo variations across the surface, but without accompanying color or IR observations, we cannot directly investigate this possibility. However, we are aware of three independent optical spectra \citep{masiero17,cbet4450b,ye17} that are all in general agreement. While the specific slope of reddening is somewhat different, the variation between them is at the tenth of a magnitude level. This is much too small to explain our large amplitude as being due to color variations across `Oumuamua's surface on the assumption that these spectra were obtained randomly in rotational phase and/or were of sufficient duration to sample a substantial fraction of the rotation. Furthermore, \citet{barucci89} showed that albedo effects tend to produce more sinusoidal lightcurves. The lightcurve's shape, with (apparently) broad, rounded peaks and a sharp minimum, is in better agreement with an elongated body.

Our lightcurve segment is insufficient for deriving a rotation state, but it can be used to place certain constraints on the rotation period of `Oumuamua.  First, the lack of repetition rules out rotation periods shorter than 3 hr. Second, if we assume that `Oumuamua is a prolate spheroid with a symmetric, double-peaked lightcurve, then the rotation period is likely to be at least 5 hr (based on the 1.3 hr interval between lightcurve minimum at $\sim$5.5 UT and the earliest time, $\sim$6.8 UT, that a maximum could reasonably be inferred thereafter). Shorter rotation periods could be possible if the lightcurve is asymmetric. Because of the incomplete nature of the lightcurve segment, we cannot place any reasonable upper limit on the rotation period.

The midpoint of our lightcurve is approximately $r'_{P1} = 23.4$. Since we did not observe a clear maximum, this is only a lower limit to a true average brightness. Our midpoint appears to be at least several 0.1 mag fainter than other magnitudes reported to the Minor Planet Center for the same time period, even after accounting for bandpass and phase angle effects. We speculate that this is due, in part, to `Oumuamua's faintness, large amplitude, and shape of its lightcurve.  The broad peaks and narrow minimum make it likely that brighter portions of the lightcurve will be sampled in a random observation. Furthermore, the MPC database is primarily for astrometry, and astrometry is best constrained when the object is brightest; thus, potentially biasing reported results to brighter magnitudes. Another possible explanation is `Oumuamua's fast rate of motion which results in very long star trails. If photometric routines measuring comparison stars trailed at `Oumuamua's rate are not adjusted to use large enough apertures, some flux from the stars will be lost without a corresponding loss from the asteroid, again biasing reported results to brighter magnitudes. Finally, uncertainties may simply be quite large due to `Oumuamua's faintness, smaller telescopes being used, very bright sky background when near the moon, and color corrections (or lack thereof) during bandpass transformations.  

Using the transformation equations from the PS1 photometric system \citep{tonry12}, 16\% per 100 nm reddening reported by \citet{cbet4450b}, and the solar flux \citep{colina96}, the midpoint brightness given above converts to $V = 23.7$ mag. Assuming $G = 0.15$ in the IAU $H$,$G$ magnitude system \citep{bowell89}, this yields an absolute magnitude, $H$, of 22.8, which is considerably larger than the JPL Small-Body Database value 22.08$\pm$0.45 (for 96 observations between 2017 Oct 14--30). However, as just discussed, there are many potential reasons why snapshot observations might yield brighter measurements than ours, and our brightest measurements are consistent with the JPL number to within the uncertainties. Our $H$ implies an effective diameter of at least $\sim$90 m for an assumed albedo ($p_v$) of 0.15, or at least $\sim$180~m for $p_{v}=0.04$ \citep{harris97}. 

`Oumuamua passed within 0.25~au of the Sun only five weeks before it was discovered and we expect that any volatile material near the surface should have begun to sublimate. To look for evidence of such putative activity, we removed a median sky background derived for each image, co-registered each frame, and created median and resistant mean stacked images with a total integration time of 9000~s (Figure~\ref{fig:stack}). We measured the radial profile for the stacked image and found a shape consistent with the radial profile of `Oumuamua measured in individual 300~s exposures (Figure~\ref{fig:rad_prof}, left panel). These are consistent with the radial profiles of the field stars in our final image. The latter, in spite of being a 3~s exposure, still resulted in trailing of $\sim$0.5~pix; however, since the trailing was in only one direction, the overall FWHM is not appreciably larger than an untrailed image. The widths of these profiles are consistent with independent assessments of the image quality throughout the night which found the best conditions to have $\sim$3 pixel seeing (0.7~arcsec).

\begin{figure*}[t]
  \centering
  \includegraphics[width=120mm]{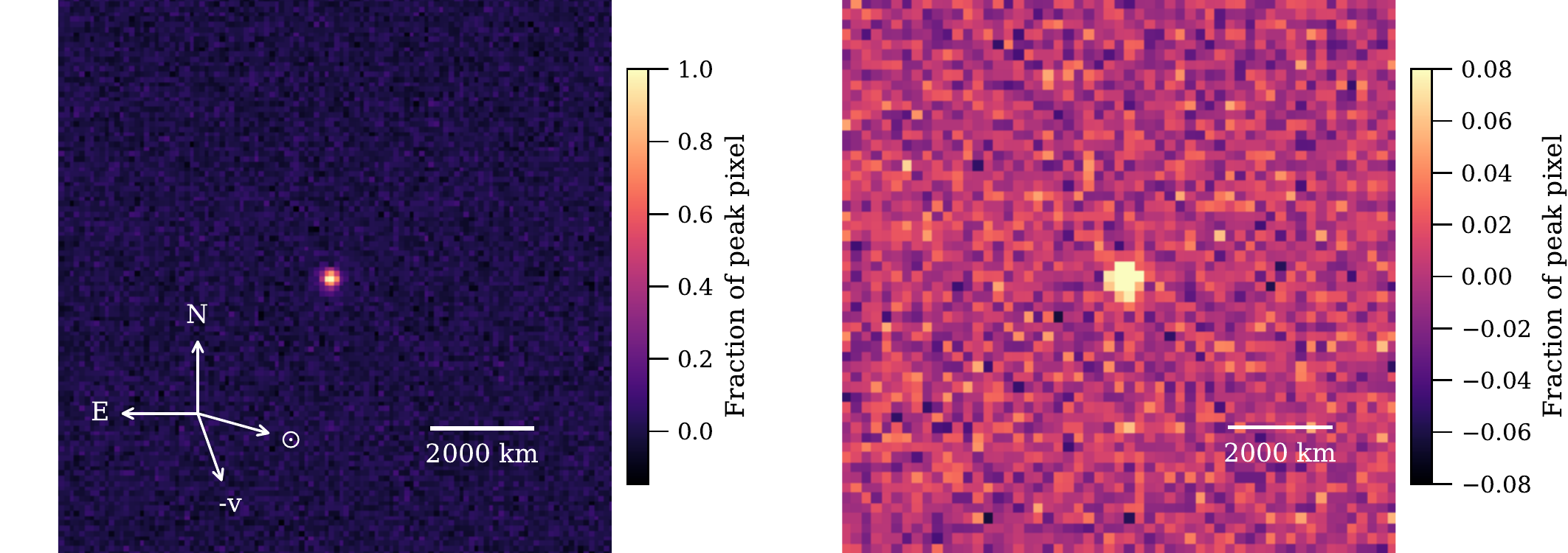}
  \caption[Stacked image]{Resistant mean stacked image with equivalent exposure time of 9000~s. No coma or tail is evident by eye. The directions to the Sun, the negative of the heliocentric velocity vector ($-v$), north, and east, as well as a scale bar are shown. {\it Left}: The image has been stretched to better show the point-source appearance of the object. {\it Right:} The image has been rebinned, 2$\times$2, and stretched to show $\pm$2$\sigma$ from the background to emphasize the lack of coma. }
  \label{fig:stack}
\end{figure*}

\begin{figure}[t]
  \centering
  \includegraphics[width=50mm]{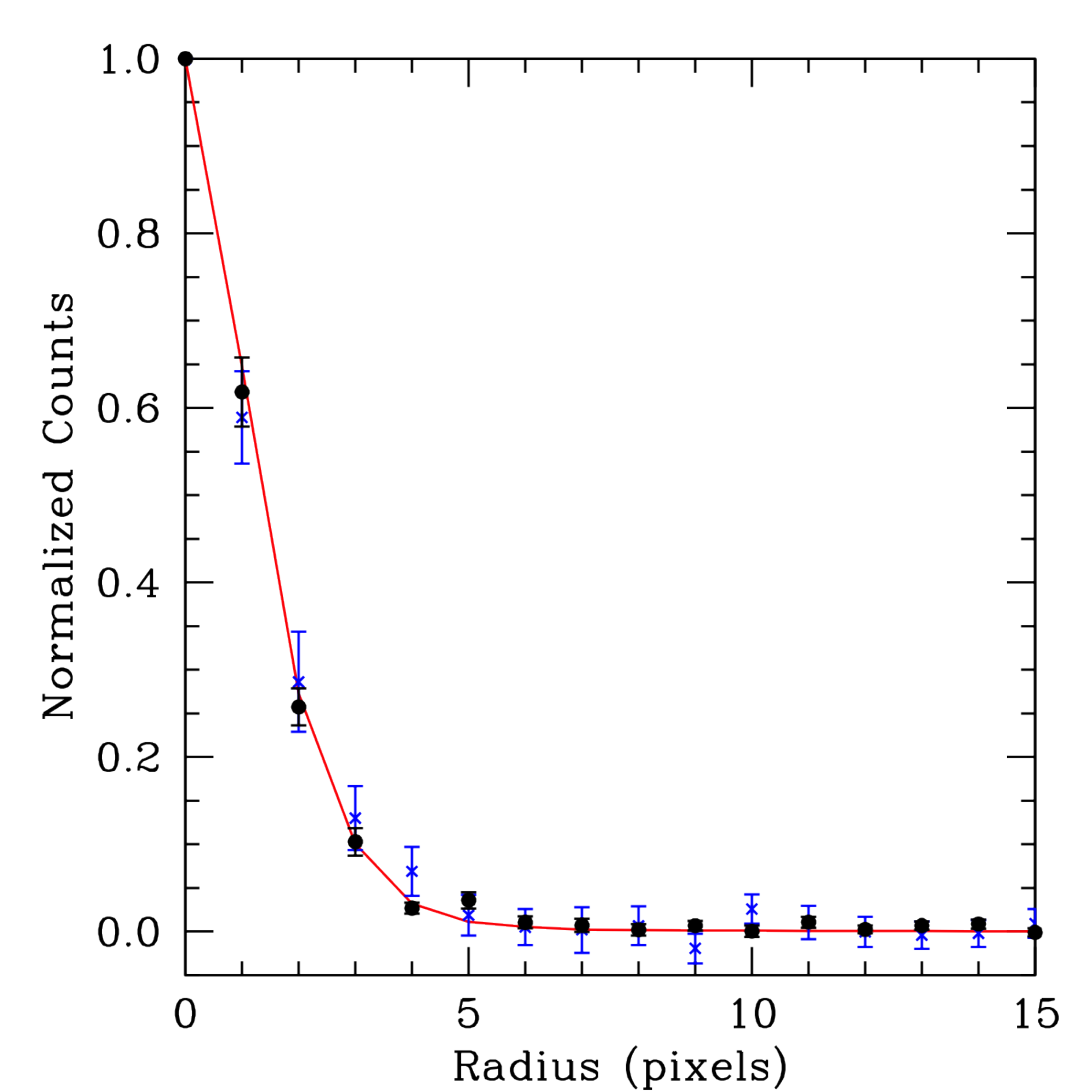}
  \includegraphics[width=50mm]{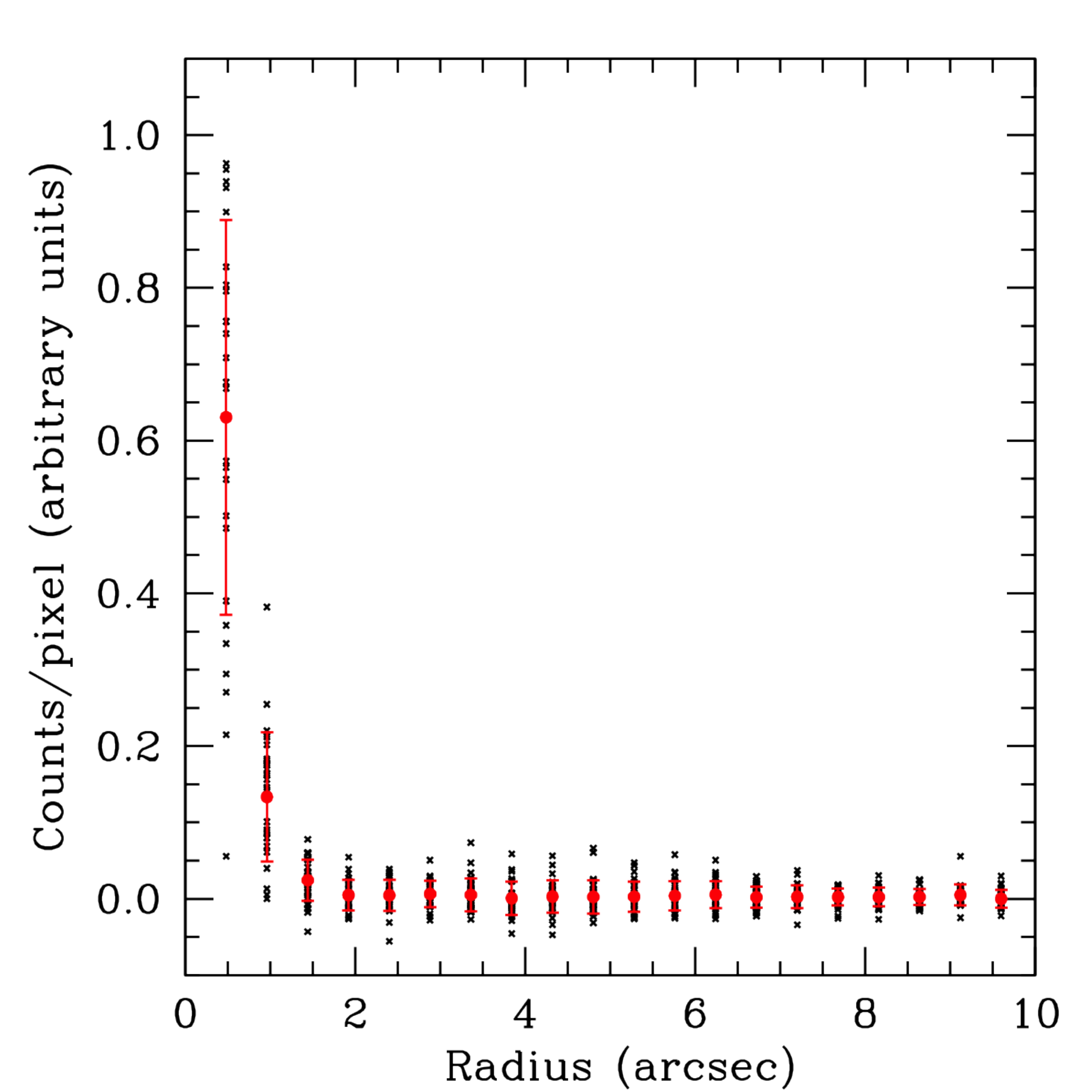}
  \caption[Nucleus properties]{{\it Top:} Normalized radial profiles of `Oumuamua in a single 300~s image (blue crosses), in the median stacked image (black circles), and of a field star in a 3~s exposure trailed at the object's rate of motion (red line). Despite the stars' slight trailing (0.5~pixel), the profiles are consistent with `Oumuamua appearing point like. {\it Bottom:} Counts per pixel in 2-pixel wide annuli. Each individual image is plotted as black crosses, and the red circles show the average and standard deviation of all 30 images.
}
  \label{fig:rad_prof}
  \label{lastfig}
\end{figure}

In order to set an upper limit to any unresolved coma that might be present, we measured the surface brightness of `Oumuamua in 2-pixel wide annuli on each image (Figure~\ref{fig:rad_prof}, right). The average and standard deviation for each annulus are overplotted in red, and the envelope of the upper error bars corresponds to a surface brightness of $\sim$27.5 magnitudes arcsec$^{-2}$ beyond 1.9 arcsec (8 pixels) in individual images. We constrained any unresolved coma in our stacked image by determining the average and standard deviation of the counts for each pixel in 2-pixel wide annuli, with the envelope of the upper error bars yielding a limiting surface brightness of $\sim$29.3 magnitudes arcsec$^{-2}$ beyond 1.9 arcsec (8 pixels). For comparison with the commonly used proxy for dust production in cometary comae, $Af{\rho}$ \citep{ahearn84}, in annuli from 3.6--4.8 arcsec (15--20 pixels) these surface brightnesses convert to upper limits for the coma $Af{\rho}$ of 0.1~cm and 0.02~cm, respectively. Typical values of $Af{\rho}$ for active comets at similar heliocentric distances are $>$1 cm (cf.\ \citealt{ahearn95}), and the comet with the smallest active area on record 209P/LINEAR had a coma $Af{\rho}$ of 0.9~cm \citep{schleicher16} when at a heliocentric distance of 1.01 au. Thus, we concur with earlier reports that `Oumuamua exhibits no evidence for cometary activity, and support its identification as observationally an ``asteroid'' with little highly volatile material. 

Measurements of `Oumuamua's lightcurve allowed us to place lower limits on its rotation period (longer than 3 hr, likely longer than 5 hr), set a lower limit to its elongation of 3:1, and infer an effective diameter of at least 90--180~m. Deep imaging ruled out the presence of any coma to a level below that of the least active comets known. These constraints are consistent with reports made public while this manuscript was under review \cite{bannister17,bolin17,jewitt17,meech17}: `Oumuamua is very highly elongated, although there is yet to be consensus on the rotation period. We are optimistic that once the ensemble of data acquired of `Oumuamua are published, our lightcurve will aide in a conclusive determination of the rotation period. Such studies will yield insight into how unique `Oumuamua is in comparison to asteroids indigenous to our own solar system.

\newpage
\section*{ACKNOWLEDGEMENTS}
We thank our telescope operator, Jason Sanborn, for assistance with observing and for acquiring dome flats. We thank the anonymous referee for a prompt review and for suggestions that greatly improved the quality of the paper. These results made use of the Discovery Channel Telescope at Lowell Observatory. Lowell is a private, non-profit institution dedicated to astrophysical research and public appreciation of astronomy and operates the DCT in partnership with Boston University, the University of Maryland, the University of Toledo, Northern Arizona University and Yale University. The Large Monolithic Imager was built by Lowell Observatory using funds provided by the National Science Foundation (AST-1005313)

This research has made use of JPL Horizons \citep{giorgini96}, SAOImage DS9, developed by the Smithsonian Astrophysical Observatory, and Astropy, a community-developed core Python package for Astronomy \citep{astropy}.

The Pan-STARRS1 Surveys (PS1) and the PS1 public science archive have been made possible through contributions by the Institute for Astronomy, the University of Hawaii, the Pan-STARRS Project Office, the Max-Planck Society and its participating institutes, the Max Planck Institute for Astronomy, Heidelberg and the Max Planck Institute for Extraterrestrial Physics, Garching, The Johns Hopkins University, Durham University, the University of Edinburgh, the Queen's University Belfast, the Harvard-Smithsonian Center for Astrophysics, the Las Cumbres Observatory Global Telescope Network Incorporated, the National Central University of Taiwan, the Space Telescope Science Institute, the National Aeronautics and Space Administration under Grant No. NNX08AR22G issued through the Planetary Science Division of the NASA Science Mission Directorate, the National Science Foundation Grant No. AST-1238877, the University of Maryland, Eotvos Lorand University (ELTE), the Los Alamos National Laboratory, and the Gordon and Betty Moore Foundation.

MMK was supported by NASA Near Earth Object Observations grant \#NNX17AK15G. SP, MSPK, and JMS thank NASA Solar System Observations grant \#NNX15AD99G for funding that supported this work.


\vspace{1mm}
{\it Facilities:} Discovery Channel Telescope (LMI)

\vspace{1mm}
{\it Software:} Astropy \citep{astropy}, IDL, SAOImage DS9
\clearpage


\end{document}